\documentclass[aps,twocolumn,showpacs]{revtex4}
\usepackage{dcolumn}
\usepackage{graphicx}
\usepackage{amsmath}
\usepackage{amsfonts}
\usepackage{amssymb}
\usepackage{psfrag}
\usepackage{wrapfig}
\usepackage{subfigure}
\usepackage{makeidx}
\usepackage{bm}
\usepackage{epsf}

\begin{document}

\title{Quantitative Relations between Modulational Instability and Several Well-known Nonlinear Excitations}
\author{Li-Chen Zhao$^{1}$}
\author{Liming Ling$^{2}$}\email{linglm@scut.edu.cn}
\address{$^{1}$Department of Physics, Northwest University, 710069, Xi'an, China}
\address{$^{2}$School of Mathematics, South China University of Technology, 510640, Guangzhou, China}
 %%%%%%%%%%%%%%%%%%%%%%%%%%%%%%%%%%%%%%%%%%%%%%%%%
\date{September 17, 2014}
\begin{abstract}
We study on the relations between modulational instability and several well-known nonlinear excitations in a nonlinear fiber, such as bright soliton, nonlinear continuous wave, Akhmediev breather, Peregrine rogue wave,  and Kuznetsov-Ma breather. We present a quantitative correspondence between them based on the dominant frequency and propagation constant of each perturbation on a continuous wave background. Especially, we find rogue wave comes from modulational instability under the ``resonance" perturbation with continuous wave background.  These results will deepen our understanding on rogue wave excitation and could be helpful for controllable nonlinear wave excitations in nonlinear fiber and other nonlinear systems.
\end{abstract}
\pacs{05.45.Yv, 02.30.Ik, 67.85.Hj, 03.70.+k}
 \maketitle

\section{Introduction}
Modulation instability (MI) is a fundamental process associated with the growth of perturbations on a continuous wave background \cite{MI}. In the initial evolution phase of MI, the spectral sidebands associated with the instability experience exponential amplification at the expense of the pump, but the subsequent dynamics are more complex and display cyclic energy exchange between multiple spectral modes \cite{Soto}.  It has many important applications in optical amplification of weak signal, material absorption and loss compensation
\cite{Abouou,Kumar}, dispersion management, all-optical switching \cite{Trillo}, frequency comb for metrology \cite{Hansson}, and so on \cite{Sylvestre, Coen, Ganapathy}.  Recently, several analytical nonlinear excitations such as Akhmediev breather(AB) \cite{AB}, Peregrine rogue wave(RW)\cite{RW},  and Kuznetsov-Ma breather(K-M) \cite{K-M}, were excited experimentally in nonlinear fiber \cite{Kibler,Dudley,Kibler2}. Even high-order RWs were excited successfully in a water wave tank \cite{Chabchoub,Chabchoub2,Chabchoub3}. The results indicate that MI can be used to understand the dynamics of these nonlinear excitations \cite{Dudley}. However, most comparisons between the properties of spontaneous MI and the analytic nonlinear excitations have been qualitative rather than quantitative \cite{RWstru}.
For example, we just know that RW should come from MI mechanism, but which mode corresponds to RW excitation has not been known precisely. The quantitative relations between MI and these nonlinear excitations can be used to find which modes are essential for RW and other nonlinear excitations. This is very meaningful for controllable nonlinear excitation in nonlinear fiber and other nonlinear systems.

In this paper, we present the quantitative relations between MI and bright soliton(BS), nonlinear continuous wave(CW), RW, AB, and K-M solutions, through defining and calculating dominant frequency and propagation constant of each perturbation signal. We demonstrate that the breathing behavior of AB or K-M comes from the frequency or propagation constant mode difference between the dominant ones of perturbation signal and CW background's. Especially, we find that RW comes from MI under the ``resonance" conditions for which the dominant frequency and propagation constant of perturbation signal are both equal to the continuous wave background's. The results would be helpful to realize controllable nonlinear wave excitations.

\begin{figure*}[t]
\centering
\subfigure[]{\includegraphics[width=85mm,height=65mm]{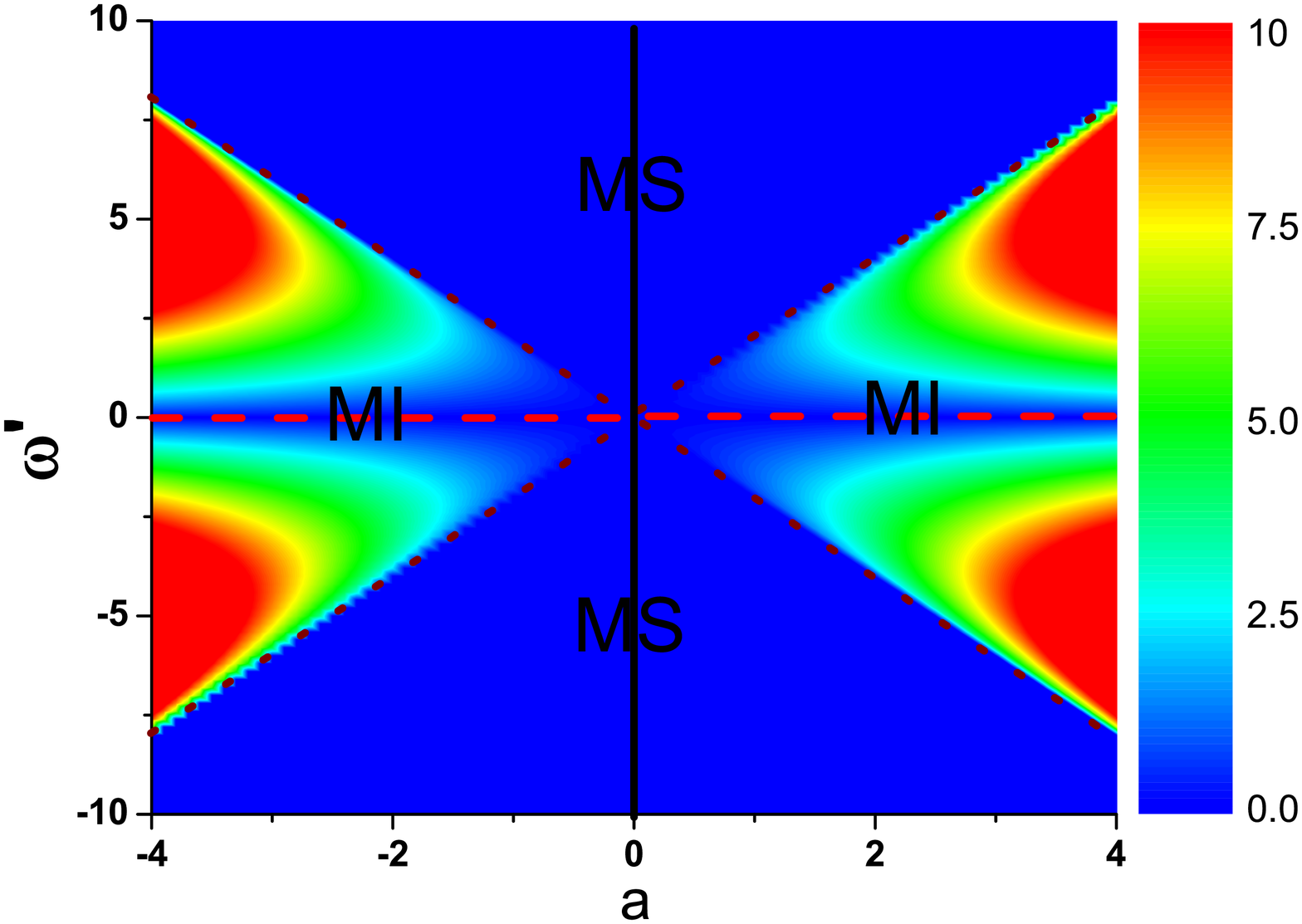}}
\hfil
\subfigure[]{\includegraphics[width=85mm,height=70mm]{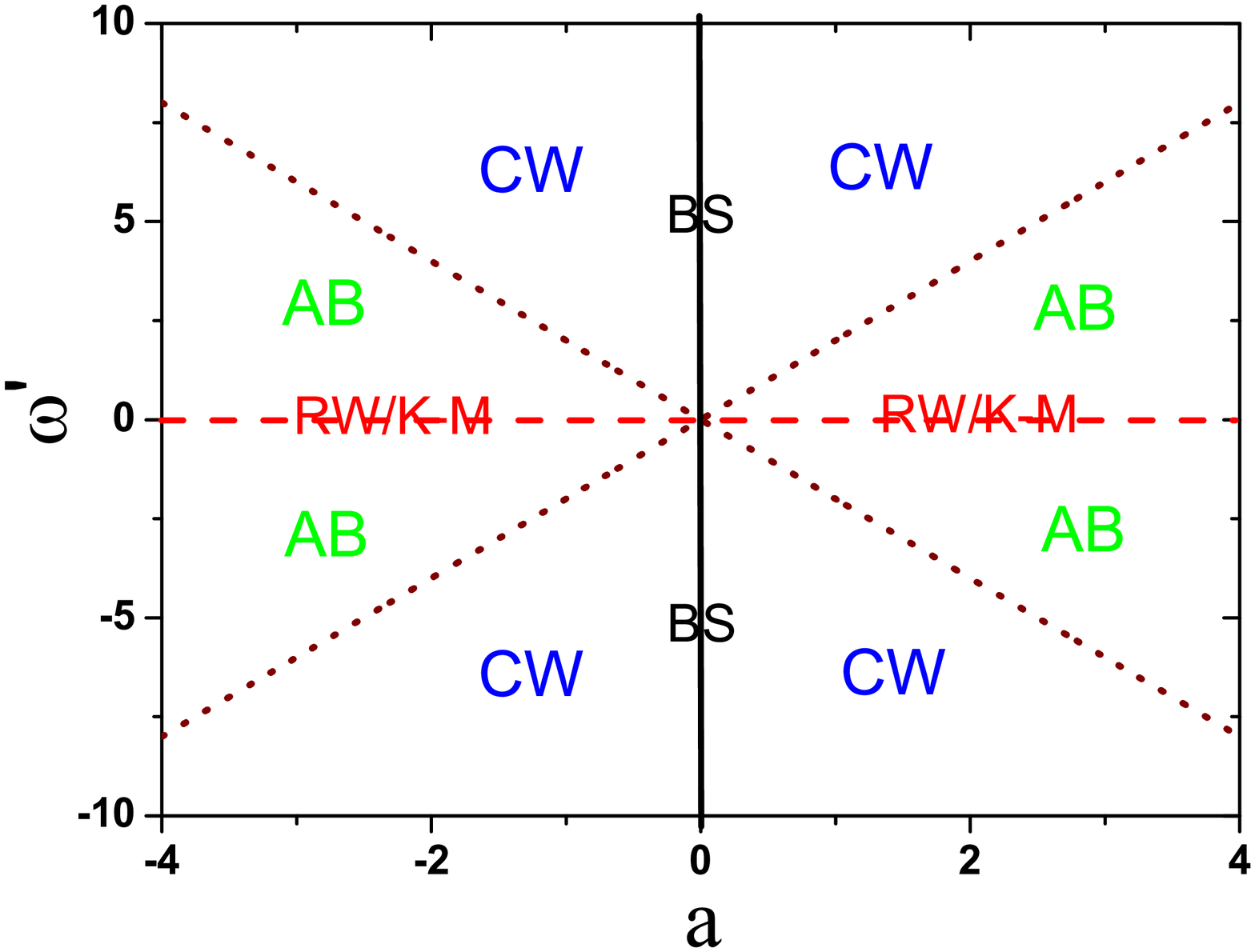}}
\caption{(Color online) (a) The modulational instability gain spectrum distributed on the continuous wave background amplitude $a$ and perturbation frequency $\omega'$ space. ``MI" and ``MS" denote modulational instability and modulational stability respectively. The red dashed line is the ``resonance line", and the black solid line is another special line for which all perturbations are stable. (b)  The  phase diagram for nonlinear waves on modulational instability gain spectrum plane. ``CW", ``AB", ``BS",
``RW", and ``K-M" denote nonlinear continuous wave, Akhmediev breather, bright soliton, Peregrine rogue wave, and Kuznetsov-Ma breather respectively. These well-known nonlinear waves of NLS are all placed clearly on the  MI plane. It should be noted that RW and K-M can not be distinctive from each other on the plane, but their differences can be clarified by dominant propagation constants of them. }\label{fig1}
\end{figure*}

\section{The relations between Modulation instability and analytic nonlinear excitations}
For simplicity without losing generality, we choose a Kerr nonlinear fiber without considering high-order effects to discuss the relations between MI and several well-known nonlinear excitations. The results can be extended to other nonlinear systems similarly. In a Kerr nonlinear fiber, the propagation of optical field (pulse duration $> 5\ ps$) can be described by the following nonlinear Schr\"{o}dinger
equation (NLSE) under slowly varying envelope approximation
 \begin{eqnarray}
i \Psi_{z}+\frac{1}{2}\Psi_{tt}+\sigma |\Psi|^2 \Psi=0.
\end{eqnarray}
The equation is written in dimensionless form \cite{Agrawal}. When the nonlinear coefficient $\sigma<0$ (corresponding self-defocusing nonlinear fiber), it admits no MI regime on CW background for which dark soliton has been found on CW background \cite{dark}.   When the nonlinear coefficient $\sigma>0$ (corresponding self-focusing nonlinear fiber),  it admits MI and modulation stability(MS) regime on the MI gain spectrum distribution. Moreover, different types of nonlinear excitations mainly including BS, AB, K-M, and RW have been derived separately \cite{Matveev,Ablo,Calini,AB,RW,K-M}, and even high-order ones \cite{Ling,He,Zhao}.  Here we study the NLSE with $\sigma>0$ to get the quantitative relations between MI and these well-known nonlinear excitations. It is convenient to set $\sigma=1$ without losing any nontrivial properties, since there is a trivial scale transformation for different values under $\sigma>0$.  Firstly, we study the MI property of NLSE  to discuss the quantitative relations.

It has been known widely that MI analysis can present us an approximate  characterization on the stability of perturbations on CW background \cite{Kuznetsov,MI}. Linear stability analysis can present us a description of MI on the amplification of each spectral mode \cite{Nithyanandan,Hansson}. We perform the standard linear instability analyze on CW background $\Psi_{0}=a \exp{[i a^2 z]}$ in the system, namely, add small-amplitude Fourier modes on the continuous wave background as $\Psi=\Psi_{0} (1+f_{+} \exp{[i\ \omega' (t-\Omega z)]}+f_{-}^* \exp{[-i\ \omega' (t-\Omega^* z)]})$ (where $f_+$, $f_-$ are small amplitudes of the Fourier modes) \cite{Wright}. Substituting them to Eq. (1) and after linearizing the equations, one can get the following dispersion relation  $\Omega'=\omega'\Omega$ where $\Omega=\frac{1}{2} \sqrt{\omega'^2-4 s^2}$. However, the MI form $\Omega'=\omega' \Omega$ can not demonstrate the MI gain value reasonably for the perturbation frequencies near the line $\omega'=0$ , since the factor $\omega'$  must be removed from the dispersion relation on the perturbation frequency $\omega'=0$ \cite{Forest}. For $\omega'=0$ mode perturbation denoted by $\tilde{p}$, we can derive the secular solution as  $\tilde{p}=1+ i\ 2 a^2 z$ which demonstrates the instability property of the resonant perturbation mode.

We demonstrate the MI gain $Im(\Omega')$ on the perturbation frequency $\omega'$ vs amplitude of the background $s$ in Fig. 1(a). It is seen that there are two distinctive regimes, namely, MI and MS. We can qualitatively know that small perturbations in MI regimes are unstable and can be amplified exponentially, and the ones in MS regime are stable and do no grow up. Especially,  there is a special line ($\omega'=0$) in the MI regime, which corresponds to the perturbation frequency equal to the CW's (writing the perturbations on CW as $\Psi=\Psi_0+f_{pert}$ form). Therefore, it can be called a ``resonant line" (red dashed line in Fig. 1(a)). It should be noted that the perturbations on this resonant line admit rational type amplification. The MI gain form $Im(\Omega')$ fails to describe the stability properties of perturbations on this line.  There is another special line $a=0$ in MS regime (black solid line in Fig. 1 (a)), on which any perturbations are all stable. The perturbations in different regimes should demonstrate different dynamical behaviors. Recent numerical simulations showed that these different dynamical processes can be described well by related nonlinear wave solutions on CW background \cite{RWstru}. However, it is hard to know which modes correspond to each type nonlinear wave excitations, since the noise include many uncontrollable different spectral modes. Notably, the analytical solutions for these nonlinear excitations can be written in the form of nonlinear continuous wave plus a perturbation term. This provides possibilities to compare them with the MI analysis on a continuous wave background, even though the MI analysis hold approximatively and nonlinear wave solutions are exact ones. It is emphasized that this idea can be used to discuss relations between MI and nonlinear excitation for many other different nonlinear systems.

 Explicitly, one term is the CW background, and the other term corresponds to perturbation term $f_{pert}$ in the above linear instability analysis. The perturbation term for these different type nonlinear excitations can be analyzed exactly by Fourier transformation. The frequency spectrums of them indeed include many different frequencies. But there is a ``dominant" one among these various frequencies, and the dominant one plays essential role in the perturbation evolution(see analyze on AB and K-M). This can be used to clarify the relations between MI and these nonlinear excitations. The main results are shown in Fig. 1(b). Nextly, we demonstrate how to locate the nonlinear waves on the MI gain spectrum one by one.

The well-known K-M solution \cite{K-M} on $\Psi_0$ background can be written as follows,
\begin{widetext}
\begin{eqnarray}
\Psi&=&  \left[a-2\frac{(b^2-a^2)\cos(2 b z \sqrt{b^2-a^2})+i b\sqrt{b^2-a^2} \sin(2 b z \sqrt{b^2-a^2})}{b \cosh(2 t \sqrt{b^2-a^2})-a\cos(2 b z \sqrt{b^2-a^2})}\right] e^{i a^2 z}.
\end{eqnarray}
\end{widetext}
The parameter $b$ determines the initial nonlinear localized wave's
shape, and $a$ is the background amplitude for
localized waves. Obviously, when $b=0$, the solution will become the nonlinear CW solution. In the following, we discuss properties of the nonlinear wave solution
according to the two parameters.  Performing Fourier transformation on the perturbation term which corresponds to the K-M dynamics, $F(\omega',z)= \int ^{+\infty}_{-\infty} f_{pert} \exp{[-i \omega' t]}\ d\ t$, we can derive the spectrum evolution of K-M. Then, we can know that the spectrum intensity is maximum at the frequency $\omega'=0$, which is called by ``dominant frequency". This can be used to understand why the temporal period of K-M is infinity. Compare with the linear instability analysis results, we can know that the dominant frequency of perturbation is equal to the CW's. The MI amplified rate is maximum one for certain $a$. Therefore, K-M excitation is on the ``resonance line" ($\omega'=0$) except the point $(a=0,\omega'=0 )$ on the MI plane (see red dashed line in  Fig. 1(b)).

When $a=0$ and $b\neq 0$, the K-M solution can be reduced to be a generalized BS solution \cite{Matveev} as
\begin{eqnarray}
\Psi&=& 2 b \,\mathrm{sech}(2 b t) e^{i 2 b^2 z}.
\end{eqnarray}
It is known that this corresponds to $a=0$ and $\omega'=0$ point on the MS spectrum. It has been found that BS is stable against small perturbations \cite{Agrawal}. Moreover, in a practical situation, the solitons can be subjected to many types of perturbations as they propagate inside an optical fiber. Therefore, BS can be located as a line $a=0$ on the MS spectrum (see black solid line in Fig. 1(b)).

When $|b|<|a|$, the K-M solution will become AB solution \cite{AB} as
\begin{widetext}
\begin{eqnarray}
\Psi&=&  \left[a-2\frac{(b^2-a^2)\cosh(2 b z \sqrt{a^2-b^2})-ib\sqrt{a^2-b^2} \sinh(2 b z \sqrt{a^2-b^2})}{b \cos(2 t \sqrt{a^2-b^2})-a \cosh(2 b z \sqrt{a^2-b^2})}\right] e^{i a^2 z}.
\end{eqnarray}
\end{widetext}
It is seen that the perturbation's frequency can be varied arbitrarily in the regime $|b| <|a|$. Spectral analysis of AB indicates that the dominant mode is $2 \sqrt{a^2-b^2}$, and there are some other much weaker nonzero frequency-multiplication modes of the perturbation. This can be proved by exact Fourier transformation \cite{ABSP}. The temporal period of AB is indeed determined by the difference between dominant frequency and CW background's. The numerical simulation also suggest that the perturbation with dominant frequency indeed evolve to be AB with corresponding period \cite{M. Erkintalo}. This means that our analysis based on dominant frequency is reasonable. Varying the parameter $b$, the perturbation's dominant mode can be changed, and the growth rate of breather changes correspondingly.   When $b=a$, the mode is on the resonance line, the AB excitation will become RW excitation \cite{Kibler}. Namely, the maximum peak and growth rate emerge on the resonance line. Therefore, we call it as resonance excitation. The MI amplification rate will become smaller with decreasing the value of $b$. When $b=0$, the mode will become the maximum frequencies $\omega'=\pm 2 a$ for AB excitations, they correspond to the points on the purple dotted line in Fig. 1(a) and (b). Namely, AB excitation can be located at the regimes between the purple dotted line and red dashed line in the MI regime.

It has been demonstrated that the spectral dynamics of MI can be described exactly by AB solution \cite{Hammani}. Based on the this, we can know that the weak perturbations with high frequencies are in the zero MI band, and there is no significant excitations on CW background. The state of system can be still seen as CW. Therefore, we denote these regimes as CW phase on MI gain distribution (see Fig. 1(b)). This agrees well with resonant vibration in classical mechanics for which large vibrations can not emerge when driving frequency is far from resonant one.

When $|b|=|a|$, the K-M and AB solution can be both reduced to RW solution \cite{RW} as
\begin{eqnarray}
\Psi&=& a \left[1-\frac{4(2ia^2z+1)}{4 a^4 z^2+4 a^2 t^2+1} \right]e^{i a^2 z}.
\end{eqnarray}
The spectrum analysis indicates that the dominant frequency of RW is on the resonant line except the point $(a=0, \omega'=0 )$ on the MI plane (red dashed line in Fig. 1(b)).  It has been shown that the resonant perturbation amplitude is indeed amplified rationally with time \cite{Forest}. Notably,  K-M and RW are both on the resonant line. Then, how to distinct RW and K-M ?

Observing the analytical expression of K-M, we can know that K-M's propagation constant is different from RW's (see $\cos(2 b z \sqrt{b^2-a^2})$ terms in Eq. (2)). With $b\neq a$ , the perturbation signal admits many propagation modes such as $2\ n b \sqrt{b^2-a^2}+a^2 $ ($n=\pm1,\pm2,\pm3,\cdot \cdot \cdot$), with writing the K-M solution as $\Psi=\Psi_0+f_{pert}$ form. The dominant one is $\pm 2 b \sqrt{b^2-a^2}+a^2$. And the oscillating period is indeed determined by the the dominant one. Therefore, the breathing behavior for K-M comes from the propagation constant difference between perturbation signal's dominant one and CW background's. With $b\rightarrow a$, the dominant perturbation propagation constant will be equal to the CW's. Namely, RW also corresponds to the resonant excitation for perturbation propagation constant.  One can prove that RW admits maximum amplification rate of small perturbations among AB, K-M and RW. This agrees well with resonant vibration in classical mechanics for which large vibrations emerge when driving frequency is close to the resonant one. It is also helpful to understand that many different forms of perturbations whose dominant frequency and propagation constant are both equal to background field's, can all evolve to be RWs \cite{Lilu}.

This could be used to understand physically the mathematical process for RW derivation, for which the spectra of Lax-pair should be degenerate for rational solution.  It is expected that it should be related with some resonance things. But this has not been known explicitly before. Here we find out that the degenerations correspond to the dominant frequency and propagation constant of the perturbation signal are both equal to the ones of CW background. This result also stands for other coupled NLS systems \cite{Zhao2, Ling2, Liu, Baronio, Baronio2, DEfoc}, and NLS with high-order effects \cite{SS,SS3, Hirota, Hirota2,Liuzhao, Wang, KE}, since the RW solutions are all derived under the degenerations of Lax-pair spectra. The rational form amplification can demonstrate larger growth rate than the exponential type ones during certain propagation distance, this makes RW be much more localized than AB and K-M \cite{RWstru}.
It should be noted that the MI analysis just can be used to understand the amplifying  process of small perturbations for nonlinear localized waves, but it can not explain the whole dynamics process of them. It is still need to develop some new ways to understand the whole dynamics process of fundamental RW and even high-order ones.

\section{Conclusion and Discussion}
We demonstrate that BS, CW, RW, AB, and K-M excitations can be located quantitatively on the MI gain spectrum plane, shown in Fig. 1(b), based on the dominant frequency and propagation constant of each perturbation form. The results are also supported by recent numerical simulations for AB and RW from many different initial conditions \cite{M. Erkintalo,Lilu}, and provide many possibilities to realize controllable nonlinear excitations. It is emphasized that the proposed method can be used to discuss relations between MI and nonlinear excitation for many other different nonlinear systems. Moreover, we find the breathing behavior of AB or K-M comes from the frequency or propagation constant mode difference between the dominant ones of perturbation signal and CW background's. Especially,  RW comes from MI under ``resonance" perturbations for which both dominant frequency and propagation constant are equal to CW background's.  This will deepen our realization on RW dynamics in many different physical systems, such as nonlinear fiber, Bose-Einstein condensate, water wave tank, plasma, and even financial systems \cite{report}.

The exact analytical solutions can not just be understood as special solutions of the nonlinear partial equation, they in fact can be used to describe fundamental dynamics of types of nonlinear excitations. This brings that they be realized experimentally even there are some deviations from the ideal initial conditions given by the exact solutions. It has been demonstrated that there are many different realistic initial conditions for AB excitations in nonlinear fiber \cite{M. Erkintalo}. Since phase and intensity modulation techniques have been developed well in nonlinear fiber, it is expected that the possibilities to realize controllable nonlinear excitations could be tested in the near future.

Note added. Recently, F. Baronio et al. suggested that RW came from the baseband MI \cite{BasM}. Explicit relations between baseband MI and resonance perturbation in MI  still need further studies.

\section*{Acknowledgments}
 This work is supported by  National Science
Foundation of China (Contact Nos. 11405129).

\end{document}